\documentclass[conference, a4paper]{IEEEtran}

\IEEEoverridecommandlockouts
\usepackage{cite}
\usepackage{amsmath,amssymb,amsfonts}
\usepackage{algorithmic}
\usepackage{graphicx}
\usepackage{textcomp}
\usepackage{xcolor}
\usepackage{listings}
\usepackage{pdfpages}
\usepackage{bm}
\usepackage{cuted}%

\usepackage{balance}%
\usepackage{footnote}
 \usepackage{etoolbox}
\makeatletter
\newcommand*{\lodbib@citeorder}{}
\newcommand*{\lodbib@notcited}{}

\usepackage[left=1.5cm,right=1.5cm,top=1.9cm, bottom=2.9cm]{geometry}
\def\BibTeX{{\rm B\kern-.05em{\sc i\kern-.025em b}\kern-.08em
    T\kern-.1667em\lower.7ex\hbox{E}\kern-.125emX}}

\usepackage{circuitikz}
\usetikzlibrary{positioning}
\usepackage{pgfkeys}

\usepackage{pgfplots}

\pgfplotsset{compat=newest}
\pgfplotsset{plot coordinates/math parser=false}
\newlength\figureheight
\newlength\figurewidth
\usetikzlibrary[patterns]
\usetikzlibrary{shapes.symbols}

\usetikzlibrary{arrows.meta,positioning}
\tikzset{block/.style={draw, rectangle, fill=cyan!90,
        minimum height=2em, minimum width=3em},
    sum/.style={draw, circle, node distance=1cm},
    input/.style={coordinate},
    output/.style={coordinate},
    pinstyle/.style={pin edge={to-,thin,black}},
        saturation block/.style={%
            draw,
            path picture={
                \pgfpointdiff{\pgfpointanchor{path picture bounding box}{south west}}%
                {\pgfpointanchor{path picture bounding box}{north east}}
                \pgfgetlastxy\x\y
                \tikzset{x=\x*.4, y=\y*.4}
                \draw [very thin] (-1,0) -- (1,0) (0,-1) -- (0,1);
                \draw [very thick] (-1,-.7) -- (-.7,-.7) -- (.7,.7) -- (1,.7);
            },
        }
    }

\tikzset{%
        rateLimit block/.style={%
            draw,
            path picture={
                \pgfpointdiff{\pgfpointanchor{path picture bounding box}{south west}}%
                {\pgfpointanchor{path picture bounding box}{north east}}
                \pgfgetlastxy\x\y
                \tikzset{x=\x*.4, y=\y*.4}
                \draw [very thin] (-1,0) -- (1,0) (0,-1) -- (0,1);
                \draw [very thick] (-1,-1) -- (1, 1);
            },
        }
    }
    \usetikzlibrary{svg.path}
    \usepackage{scalerel}

    \definecolor{orcidlogocol}{HTML}{A6CE39}
    \tikzset{
      orcidlogo/.pic={
        \fill[orcidlogocol] svg{M256,128c0,70.7-57.3,128-128,128C57.3,256,0,198.7,0,128C0,57.3,57.3,0,128,0C198.7,0,256,57.3,256,128z};
        \fill[white] svg{M86.3,186.2H70.9V79.1h15.4v48.4V186.2z}
                     svg{M108.9,79.1h41.6c39.6,0,57,28.3,57,53.6c0,27.5-21.5,53.6-56.8,53.6h-41.8V79.1z M124.3,172.4h24.5c34.9,0,42.9-26.5,42.9-39.7c0-21.5-13.7-39.7-43.7-39.7h-23.7V172.4z}
                     svg{M88.7,56.8c0,5.5-4.5,10.1-10.1,10.1c-5.6,0-10.1-4.6-10.1-10.1c0-5.6,4.5-10.1,10.1-10.1C84.2,46.7,88.7,51.3,88.7,56.8z};
      }
    }

    \newcommand\orcidicon[1]{\href{https://orcid.org/#1}{\mbox{\scalerel*{
    \begin{tikzpicture}[yscale=-1,transform shape]
    \pic{orcidlogo};
    \end{tikzpicture}
    }{|}}}}
    \usepackage[hidelinks]{hyperref}

\begin{document}
\title{Novel Joint Estimation and Decoding Metrics\\for Short-Blocklength Transmission Systems}
\author{\IEEEauthorblockN{ Mody~Sy~\IEEEmembership{}\orcidicon{0000-0003-2841-2181} ~
        and~ Raymond~Knopp~\IEEEmembership{}\orcidicon{0000-0002-6133-5651}}
        \IEEEauthorblockA{
        \texttt{EURECOM, 06410 BIOT, France} \\
        mody.sy@eurecom.fr,~raymond.knopp@eurecom.fr
        }
}
%
\maketitle
\begin{abstract}
This paper presents Bit-Interleaved Coded Modulation metrics for joint estimation detection using training or reference signal transmission strategies for short to long block length channels. We show that it is possible to enhance the performance and sensitivity through joint detection-estimation compared to standard receivers, especially when the channel state information is unknown and the density of the training dimensions is low.  The performance analysis makes use of a full 5G transmitter and receiver chains for both Polar and LDPC coded transmissions paired with BPSK/QPSK modulation schemes. We consider transmissions where reference signals are interleaved with data and both are transmitted over a small number of OFDM symbols so that near-perfect channel estimation cannot be achieved.
Our findings demonstrate that when the detection windows used in the metric units is on the order of four modulated symbols the proposed BICM metrics can be used to achieve detection performance that is close to that of a coherent receiver with perfect channel state information for both polar and LDPC coded configurations. Furthermore, we show that for transmissions with low DMRS density, a good trade-off can be achieved in terms of additional coding gain and improved channel estimation quality by adaptive DMRS power adjustment.
\end{abstract}


\section{Introduction}
It is expected that the 6G air-interface will build upon the 5G standard and address new pardigms for feedback-based cyber-physical systems combining
communications and sensing. In particular, there will be a need for tight control loops using the air-interface to control 6G-enabled objects with
high-reliability, perhaps even requiring lower latencies than those achieved by current 5G technology, for example sub-1ms uplink application-layer latency in  microwave spectrum. Although 5G transmission formats can provide very short-packet transmission through the use of mini-slots, the ratio of tranining
information to data is not necessarily adapted to extremely short data transmission. Moreover,
the transmission formats are designed with conventional quasi-coherent receivers which can be quite sub-optimal in such scenarios where accurate
channel estimation is impossible because of sporadic transmission of short packets. One such instance is because of stringent decoding latency
constraints such as those emerging in so-called {\em Ultra-Reliable-Low-Latency Communication} (URLLC) industrial IoT applications. This would be
similar for evolved channel state information (CSI) feedback control channels or future combined-sensing and communication paradigms requiring rapid sensory
feedback to the network.\\
The area of short block transmission has garnered significant attention in recent years, with extensive research conducted on various aspects, including the design of signal codes, receiver algorithms \cite{Lee2018}\cite{sy2023} and the establishment of state-of-the-art converse and achievability bounds for both coherent and non-coherent communications \cite{Xhemrishi2019,Yuan2021, Polyanskiy2010, Durisi2016, Ostman2019jrnal}.

 In this work, we investigate bit-interleaved coded modulation (BICM) and detection strategies for packets in the range of 20-100 bits for these envisaged 6G signaling scenarios.

Zehavi \cite{zehavi92} proposed bit-interleaved coded modulation (BICM) as a pragmatic approach to coded modulation. Its basic principle is the ability of an interleaving permutation to separate an underlying binary code from an arbitrary higher-order modulation \cite{Fabregas08}. Per-bit log-likelihood ratios are used to convey soft metrics from the demodulator to the decoder in order to reduce information loss. This fundamental observation spurred interest in BICM and cemented its position as a standard coding technique in wireless communication channels. In \cite{CTB98}, Caire et al. later conducted a thorough analysis of BICM in terms of information rate and error probability including both coherent and non-coherent detection. Today, BICM is widely considered as the cornerstone of high spectral efficiency systems, as well as low spectral efficiency orthogonal modulation systems. Since the 3G era, BICM has been employed in 3GPP systems. Therefore, in order to improve their efficiency and enable high-performance  communication, schemes such as rate matching, scrambling and other processes inherent in  modern wireless communication standards are de facto added to the reference BICM schematic. In addition, the underlying detection and decoding metrics must offer enhanced performance and low complexity trade-off.
Thus we examine BICM metrics exploiting joint detection and estimation which are amenable to situations where low-density demodulation reference signals  are interleaved with coded data symbols. We consider standard OFDM transmission so that both DMRS and data are interleaved in frequency.  We show that by using a properly conceived metric exploiting interleaved DMRS in the decoding metric computation, we can achieve performance approaching a receiver with perfect channel estimation and significant coding gains compared to a conventional 5G OFDM receiver. The scheme performs detection over contiguous groups of modulated symbols including those from the DMRS to provide soft metrics for the bits in each group to the channel decoder. We evaluate performance using a full 5G transceiver chain for both polar and LDPC coded formats with up to eight receive antennas. The schemes are applicable to both uplink and downlink transmission where packets are encoded into a small number of OFDM symbols with interleaved DMRS. Additionally, we investigate the impact of varying densities of reference signals on performance.
The remainder of this article  is structured as follows. Section II lays out the system model and foundations of NR polar and LDPC-coded modulations, Section III highlights the proposed BICM Metrics, Section IV presents the results and performance analysis, and finally Section V concludes the paper.\\
\emph{Notation :}
Scalars are denoted by italic letters, vectors and matrices are denoted by bold-face lower-case and upper-case letters, respectively.
 For a complex-valued vector $\mathbf x$, $|\lvert \mathbf x |\rvert$ denotes its Euclidean norm, $| \cdot |$  denotes the absolute value.  
   $ \mathbb E\{\cdot\}$ denotes the statistical expectation. $\operatorname{Re}({\cdot})$ denotes the real part of a complex number. $\operatorname{I_0}(\cdot)$ is the zero-th order modified Bessel function of the first kind.
$\mathbf I$ is  an identity matrix with appropriate dimensions.
Galois field  is denoted by $GF(2)$ or $\mathbb F_2$.
 $\mathbf x \in \chi^j_b=\left\{\mathbf x: e_j=b\right\}$ is the subset of symbols $\{\mathbf x\}$ for which the $j-th$  bit of the label $e$ is  equal to $b=\{0,1\}$.  At the $j-th$ bit location or position and the number of bits reqired to a symbol is denoted by $m\triangleq \log_2\left(M\right)$.
 The cardinality of $\chi$ is given by $M\triangleq |\chi|$.
$ \Lambda^j\left(\cdot\right)$ denotes log likelihood ratio, with $j=1,2, \ldots, m$.
The superscripts  $^T$  and $^*$ or $^\dag$ denote the transpose and  the complex conjugate transpose or Hermitian.
\section{Preliminaries}
\subsection{System Model}
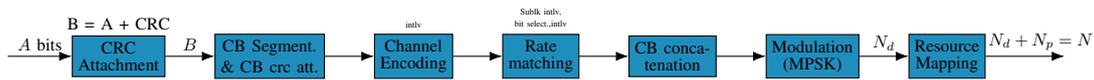
\begin{figure*}[ht]\centering
      \begin{tikzpicture}[auto, node distance=0.5cm,>=Latex, scale=.65, transform shape]
      \node [input, name=input] {};
      \node [block, fill={rgb, 255:red, 32; green, 142; blue, 189 }, right=1.3cm of input, name = sum,minimum height=1em, minimum
      width=4em, label = { B = A + CRC}] {\shortstack{CRC\\ Attachment}};
      \node [block,fill={rgb, 255:red, 32; green, 142; blue, 189 }, right=1cm of sum, label = {}, minimum height=1em, minimum
      width=3em] (ctrler) {\shortstack{CB Segment. \\ \& CB crc att.}};
      \node [block,fill={rgb, 255:red, 32; green, 142; blue, 189 }, label=above:\tiny{intlv}, minimum height=1em, minimum
      width=4em, right=1cm of ctrler] (sat){\shortstack{Channel\\Encoding }};
      \node [block,fill={rgb, 255:red, 32; green, 142; blue, 189 }, right=1cm of sat, label=above:\tiny{\shortstack{Sublk intlv, \\bit select.,intlv }}, minimum height=1em, minimum
        width=3em] (rate){\shortstack{Rate \\ matching }};
    \node [block,fill={rgb, 255:red, 32; green, 142; blue, 189 }, label=above:\tiny{}, minimum height=1em, minimum
      width=4em, right=1cm of rate] (concat){\shortstack{CB conca-\\tenation }};
    \node [block,fill={rgb, 255:red, 32; green, 142; blue, 189 }, right=1cm of concat, label = {}, minimum height=1em, minimum
      width=3em] (mod) {\shortstack{Modulation \\(MPSK) }};
     \node [block,fill={rgb, 255:red, 32; green, 142; blue, 189 }, right=1cm of mod, label = above:\tiny{}, minimum height=1em, minimum
      width=3em] (pucch) {\shortstack{Resource \\Mapping }};
    \node [output, right=1.6 of pucch, node distance = 2 cm] (output) {};
      width=3em, right=1cm of pucch] (rateLim) {DAC \& RF };
      \draw [draw,->] (input) -- node[pos=0.5] {} node {$A$ bits} (sum);
      \draw [->] (sum) -- node [name = comm, pos = 0.5]{$ B $}(ctrler);
      \draw [->] (ctrler) -- node [name = commSig, pos = 0.5]{}(sat);
      \draw [->] (sat) -- node [name = contSig1, pos = 0.5]{ } (rate);
      \draw [->] (rate) -- node [name = contSig1, pos = 0.5]{ } (concat);
      \draw [->] (concat) -- node [name = contSig, pos = 0.5]{} (mod);
     \draw [->] (mod) -- node [name = contSig, pos = 0.5]{$N_d$} (pucch);
      \draw [->] (pucch) -- node [name = contSig, pos = 0.7]{$N_d+N_p=N$} (output);
      \end{tikzpicture}
      \centering
      \caption{Bit-Interleaved Polar/LDPC coded Modulation (BIPCM/BILCM) : Transmitter end.}
      \label{fig:bicm_polar_ldpc}
\end{figure*}

Consider a SIMO OFDM BICM system with a single antenna element on the transmit array and multiple element receive arrays. The system dimensions are defined as $(N_R \times 1)$, where $N_T =1$ and $N_R$ refer to the number of antennas on the transmitter and receiver, respectively. The transmitted and received signals are $N$-dimensional column vectors, and thus a system is designed in such a way that the relationship between the transmitted and received signals is as follows:
\begin{equation} \label{eqn:systmodel}
\centering
    \mathbf{y}_i = \mathbf h_i\mathbf{x}+ \mathbf{z}_i, \quad i=0,1,\cdots,N_R-1,
\end{equation}
where $\mathbf{y}_i$  represents an observed vector in $N$ complex dimensions, $\mathbf{x}$  is  an $N$-dimensional modulated vector transporting $B$ channel bits, so that the message $m = 0,1 \ldots, 2^B-1$, $z$ is additive white Gaussian noise whose  real and imaginary components are independent and have variance $\sigma^2$ in each dimension. Various models for $\mathbf h$ will be used in this study and will be described along with the corresponding receiver structures.
The transmitted vector $\mathbf{x}$ is often composed of data independent components which are known to the receiver. These are so-called {\em pilot} or {\em demodulation reference signals (DMRS)} which are conceived in order to allow for resolving channel ambiguity in time, frequency and space. In practice, the reference  signals are used for estimating  the vector channels $\left\{h_i\right\}$ and are commonly interleaved among the data-dependent components according the characteristics of the propagation channel. It is notably the case in current OFDM systems. In earlier CDMA systems, DMRS were sometimes superimposed on top of data-dependent signals. We denote the number of data dimensions by $N_d$ and reference signal dimensions by $N_p$ where $N_d+N_p=N$.
The assumption in this work is that the data-dependent components of $\mathbf{x}$ are generated from a binary code whose output is interleaved and mapped to an $M$-ary modulation symbol alphabet. We will assume that the binary code generates $E$ bits and the interleaver mapping is one-to-one so that $E$ bits are also fed to the modulator. The binary-code and interleaver combination can thus be seen as a $(E,B)$ binary block code. Denote the $E$ coded bits as $e_k,k=0,1,\cdots, E-1$. Adjacent $\log_2 M$ bit-tuples are used to select the $N_E$ modulated symbols in the symbol alphabet. Typically, we will assume that a Gray mapping is used in the case of non-binary modulation.\\
Bit Interleaved Polar Coded Modulation is referred to as BIPCM in this paper and makes use of  CRC-Aided Polar (CA-Polar) codes, one of the basic code construction techniques established by the 3GPP Standard\cite{3GPP38212}.
In addition, Bit Interleaved LDPC Coded Modulation is referred to as BILCM. 
The overall representation of the BIPCM/BILCM schematic is shown in Figure~\ref{fig:bicm_polar_ldpc}. This figure depicts the transmit-end procedure for uplink channels.

\begin{figure}[!ht]
        \centering
        \input{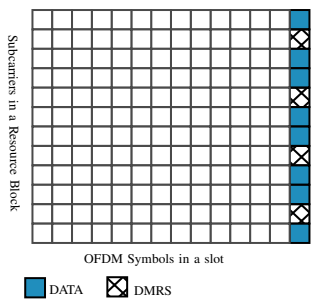}
        \caption{General resource mapping: 1 OFDM  symbol.}
        \label{fig:re_mapp_pucch2_1symb}
\end{figure}

In both scenarios, the encoded payload undergoes rate-matching and block concatenation prior to being fed to a QPSK modulator. This process yields a set of complex-valued modulation symbols, represented as $x(0), x(1), \ldots, x\left(N_d / 2-1\right)$. Subsequently, the resource allocation process is executed, wherein one or multiple OFDM symbols are utilized to allocate the modulated symbols to resource blocks and insert the DMRS resources.
As illustrated in  Figure~\ref{fig:re_mapp_pucch2_1symb}, the resource mapping  here is embedded in the same spirit as in a 3GPP PUCCH2 transmission.

\subsection{Perfect Channel State Information}
We denote the likelihood function for the observed vector on a particular receiver branch  as
 \begin{equation}\label{eqn:lfpcsi1}
 \begin{aligned}
 q\left(\mathbf{x},\left\{\mathbf{y}_i, \mathbf h_i\right\}\right)=p\left(\left\{\mathbf{y}_i, \mathbf h_i\right\} \mid \mathbf{x}\right)=p\left(\left\{\mathbf{y}_i\right\} \mid \mathbf{x}, \{\mathbf h_i\}\right) p\left(\left\{\mathbf h_i\right\} \mid \mathbf{x}\right)
\end{aligned}
 \end{equation}
 If the transmitted signal ${\mathbf x}$ is independent of the channel realization $\{\mathbf h_i\}$,  the term $p\left(\left\{{\mathbf h}_i\right\}\mid\mathbf{x}\right)$ in (\ref{eqn:lfpcsi1}) can be dropped. The likelihood function is equivalent to $ q\left(\mathbf{x},\left\{\mathbf{y}_i, \mathbf h_i\right\}\right)=\frac{1}{(\pi N_0)^N} \exp \left(-\frac{\lvert|\mathbf y_i-\mathbf h_i \mathbf x\rvert|^2}{N_0}\right)$.
Using the norm extension property, ignoring terms that are independent of $\mathbf x$,
 then the likelihood function is
 \begin{equation}\label{eqn:lfpcsi3}
 \begin{aligned}
 &q\left(\mathbf{x},\left\{\mathbf{y}, \mathbf{h}\right\}\right) \propto
 &\prod_{i=0}^{N_R-1}\exp \left(\frac{ 2\operatorname{Re}\left(\mathbf{y}_i \mathbf h_i^* \mathbf{x}^*\right) -\lvert|\mathbf h_i \mathbf{x}\rvert|^2}{N_0}\right).
 \end{aligned}
 \end{equation}
 The likelihood of coded bit $e_j\in\{0,1\}$ is
 \begin{equation}\label{eqn:lf_simo}
 q_{j,b}\left(\mathbf y_i\right)=\sum_{\mathbf x \ \in \ \chi_b^j}q\left(\mathbf{x}, \mathbf y_i\right).
 \end{equation}

 As is common in the case of BICM-based systems, the soft input to the binary channel decoder is given as the log-likelihood ratio (LLR) for coded bit $j$.
 \begin{equation}\label{eqn:llr_simo}
 \Lambda^j\left(\mathbf y_i\right)=\log\frac{q_{j,0}\left(\mathbf y_i\right)}{q_{j,1}\left(\mathbf y_i\right)}.
 \end{equation}



 We typically simplify (\ref{eqn:llr_simo}) via a \emph{max-log approximation}
 $\log \left\{\sum_{i} \exp \left(\lambda_{i}\right)\right\} \sim \max _{i}\left\{\lambda_{i}\right\}$ letting (\ref{eqn:llr_simo}) to be simplified as
  \begin{equation}\label{eqn:maxlog_simo_conv_rx}
  \begin{aligned}
 \Lambda^j\left(\mathbf y\right)&=  \displaystyle \max _{\mathbf x \ \in \ \chi_0^j}\frac{1}{N_0} \sum_{i=0}^{N_R-1}2\operatorname{Re}\left(\mathbf{y}_i \mathbf h_i^* \mathbf{x}^*\right) -\lvert|\mathbf h_i \mathbf{x}\rvert|^2 \\&-\displaystyle \max _{\mathbf x \ \in \ \chi_1^j} \frac{1}{N_0} \sum_{i=0}^{N_R-1}2\operatorname{Re}\left(\mathbf{y}_i \mathbf h_i^* \mathbf{x}^*\right) -\lvert| \mathbf h_i \mathbf{x}\rvert|^2.
  \end{aligned}
  \end{equation}
 This metric is typically used in {\em Perfect CSI} based receivers as well as in conventional quasi-coherent receivers employing separate least-squares channel estimation by substituting $\mathbf{h}$, with  $\hat{\mathbf h}$.
We consider the ideal or {\em Perfect CSI} receiver as a benchmark for comparison with the subsequent receiver architectures.

Moreover, within the framework of a conventional receiver, it is presupposed that, at the very least, the observation of a single reference signal spans a PRB in order to generate the coded bits corresponding to each data symbol in that PRB.
In addition, this conventional metric applies a so-called {\em symbol-by-symbol} detection, that’s to say that each symbol is detected independently of the other symbols.
We will note this as {\em No CSI} ($N_d=1$).
\section{BICM Receivers}
  \subsection{ Metrics for Non-Coherent Fading Channels}
  We describe BICM metrics for a general non-coherent fading channel with unknown phase on the line-of-sight (LOS) components
  and fully unknown diffuse (Non-LOS) components. The overall unknown channel gain is given by
  $\mathbf {h}_i=\left(\sqrt{\alpha}e^{j\theta_i}+\sqrt{1-\alpha}\mathrm h_i^{(f)}\right)\mathbf{I}$ where $\theta_i$ is assumed to
  be i.i.d. uniform random variables on $[0,2\pi)$, $\mathrm h_i^{(f)}$ is a zero-mean
  unit-variance circularly-symmetric complex Gaussian random variable and $\alpha$ is the relative strength of the LOS component.
  The amplitude $|\mathrm h_i|$ on each receiver is thus Ricean distributed. It is worth noting that
  the i.i.d. assumption for the ${\theta_i}$ is somewhat unrealistic for a modern array receiver with accurate calibration.
  The phase differences would be more appropriately characterized by two random-phases, one originating from the time-delay between transmitter and receiver and the other from the angle of arrival of the incoming wave. The phase differences of individual
  antenna elements for a given carrier frequency could then be determined from the angle of arrival and the particular
  geometry of the array. To avoid assuming a particular array geometry, the i.i.d. uniform model provides a simpler and universal means to derive a receiver metric.\\
  \textbf{Propositionn 1:}
   The corresponding likelihood function after neglecting multiplicative terms independent of the transmitted message, can be shown to be
  \begin{equation}\label{eqn:llrfunction_simplified}
  \begin{aligned}
  q\left(\mathbf{x},\mathbf y\right)&= \prod_{i=0}^{N_R-1} \frac{1}{\mathbf L_x}\exp \left(  -\frac{ \alpha \left\|\mathbf x \right\|^2 }{\mathbf L_x}  + \right.\\&\left. \beta_x \left| \mathbf x^\dag \mathbf y_i\right|^2\right)\times\operatorname{I_0}\left(  \frac{2\sqrt{\alpha}}{\mathbf L_x}\left|\mathbf x^\dag\mathbf y_i\right|\right)
  \end{aligned}
  \end{equation}
  where
  $\mathbf L_x = N_0+ 2(1-\alpha) \left\|\mathbf x \right\|^2$,  $\mathbf \beta_x = \frac{2(1-\alpha)}{N_0(N_0 +2(1-\alpha) \left\|\mathbf x \right\|^2)}$ and $\operatorname{I_0}(\cdot)$ is the zero-order modified Bessel function.\\
Note that in the resulting expressions  of  LLR of coded bit, we do not limit the dimensionality of the observations when computing likelihoods of particular bits. In the original work of Caire {\em et al}\cite{CTB98} the authors assume an ideal interleaving model which allows limiting the observation interval of a particular coded bit to the symbol in which it is conveyed. For long blocks this assumption is realistic for arbitrary modulation signal sets and is sufficient for BPSK and QPSK irrespective of the block length when the channel is known perfectly. Nevertheless, practical systems usually apply single symbol likelihood functions for short blocks and high-order modulations.
  \subsubsection*{Proof}
  See appendix section.\\
  \textbf{Corollary:}
  Metric calculations based on  (\ref{eqn:llrfunction_simplified}) are computationally complex from an implementation perspective and
  are typically simplified. As is the case for
  the known channel, we can apply the \emph{max-log approximation} after first using an exponential approximation $I_{0}(z) \sim \frac{e^{z}}{\sqrt{2 \pi z}} \sim e^{z}$
  yielding the approximated log-likelihood ratio (LLR) for coded bit $j$.
  \textbf{Remark:}
  Note that in (\ref{eqn:maxlog_llr_simo}), many of the terms can be dropped when $\|\mathbf{x}\|$ is constant,
  as it would be the case for BPSK or QPSK modulation for instance. Strong LOS channels can also neglect the quadratic terms in (\ref{eqn:maxlog_llr_simo}).

  When $\alpha=1$, (\ref{eqn:maxlog_llr_simo}) corresponds to a pure LOS channel.
  \subsection{Joint Estimation and Detection Principle}
  For the case of polar or LDPC-coded data, there is a compelling motivation to divide the coded streams into smaller blocks for detection due to complexity reasons. Assuming an ideal interleaving scenario with known channels \cite{CTB98}, detection can be performed on individual modulated symbols. However, in the presence of joint detection and estimation, where interleaved DMRS  and data symbols are considered, we need to deal with short blocks that encompass both data and DMRS symbols. To achieve this, the $N$-dimensional vectors $\mathbf{y}$ and $\mathbf{x}$ are subdivided into smaller segments of $N_b$-dimensional blocks. Subsequently, the bit LLR (Log-Likelihood Ratio) metric is applied to each of these underlying segments for further processing and analysis.\\
  \textbf{Proposition 2:}
  Observing the structure of the metrics and the absence of overlap between the data and DMRS symbols,
  we can easily see that the estimated channel impulse response (CIR)    is part of the metrics.
    By writing $\mathbf x = \mathbf x^{(d)} + \mathbf x^{(p)} $
    where  $d$ and $p$ are subscripts representing data, DMRS components, respectively, we can
    reveal $\hat{\mathbf h}^{\mathrm{LS}}_i$ in the metrics:
    \begin{align}
    \left|{\mathbf x}^\dag{\mathbf y_i}\right| &= \left|\underbrace{{\mathbf x^{(p)}}^\dag\mathbf{y}^{(p)}_{i}}_{\text {channel estimate}}  + \ {\mathbf x^{(d)}}^\dag\mathbf{y}^{(d)}_{i}\right|.
    \label{eqn:joint_estim_simo}
    \end{align}

    The estimation of the channel's characteristics is achievable through the computing of the correlation between the reference transmitted signal and the reference received signal.

    Mathematically, the channel estimate  can be obtained via the joint  least squares (LS) method as follow ${\mathbf x^{(p)}}^\dag{\mathbf{y}^{(p)}_i}=\left({\mathbf x^{(p)}}^\dag \mathbf x^{(p)}\right) \hat{\mathbf h}_i^{{\text{\tiny{LS}}}}= \|\mathbf x^{(p)}\|^2\cdot\hat{\mathbf h}_i^{{\text{\tiny{LS}}}}=N_p \cdot \rho \cdot\hat{\mathbf h}_i^{{\text{\tiny{LS}}}}$
    where  $N_p$ number of pilots, $\rho$ is the reference signal  power and is typically normalized to unity.
    Then, equation (\ref{eqn:joint_estim_simo}) is equivalent to :
     \begin{align}
     \left|{\mathbf x}^\dag{\mathbf y_i}\right| &=\left|N_p\hat{\mathbf h}^{{\text{\tiny{LS}}}}_i +  \ {\mathbf x^{(d)}}^\dag\mathbf{y}^{(d)}_{i}\right|.
     \label{eqn:joint_estim_simo_simplified}
     \end{align}
     where $\hat{\mathbf h}^{{\text{\tiny{LS}}}}$ is the channel impulse response obtained via a joint  least-squares (LS) channel estimation using averaging or smoothing
    over an appropriate number of dimensions exhibiting channel coherence. In the process of short-block detection, we can make use of such a channel estimate that. In general, the channel estimation procedure will work as usual and the resulting estimates are fed into the metrics considered here.\\
    \begin{strip}
    \hrulefill
    \begin{align}\label{eqn:maxlog_llr_simo}
    \begin{array}{r}
    \Lambda^j\left(\mathbf y\right)= \displaystyle\max _{\mathbf x \ \in \ \chi_0^j}\left( \displaystyle\sum_{i=0}^{N_R-1}-\frac{\alpha\left\|\mathbf{x}\right\|^2}{\mathbf{L}_x}+\mathbf{\beta}_x\left|\mathbf{x}^\dag \mathbf{y}_i\right|^2+\frac{2 \sqrt{\alpha}}{\mathbf{L}_x}\left|\mathbf{x}^\dag\mathbf{y}_i\right|\right)-\sum_{\mathbf x \ \in \ \chi_0^j} \displaystyle N_R\log \left(\mathbf{L}_x\right) \\
    \quad- \displaystyle\max _{\mathbf x \ \in \ \chi_1^j}\left( \displaystyle\sum_{i=0}^{N_R-1}-\frac{\alpha\left\|\mathbf{x}\right\|^2}{\mathbf{L}_x}+\mathbf{\beta}_x\left|\mathbf{x}^\dag \mathbf{y}_i\right|^2+\frac{2 \sqrt{\alpha} }{\mathbf{L}_x}\left|\mathbf{x}^\dag\mathbf{y}_i\right|\right)+  \displaystyle\sum_{\mathbf x \ \in \ \chi_1^j}N_R\log\left(\mathbf{L}_x\right).
    \end{array}
    \end{align}
    \hrulefill
    \end{strip}
\section{Numerical Results}
\subsection{Metric Performance Analysis}
\begin{figure*}[!ht]
\begin{minipage}[c]{.32\linewidth}
  \centering
  \includegraphics[width=\linewidth]{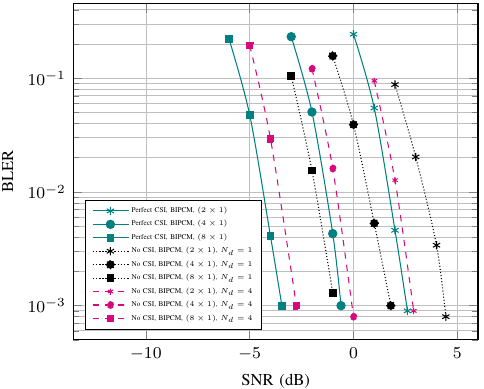}
  \caption{Block Error Rate, 48 bits(TBs+CRC), NRPOLAR BICM, R=48/64, (CRC)-aided successive-cancellation list decoder(List length=8), QPSK modulation, 1 OFDM symbol, 4 PRBs, 48 REs (32 data, 16 DMRS), $N_T$=1, $N_R = \{2, 4, 8\}$, $\alpha=1$, LOS channel. }
  \label{fig:bicm_pucch48bits_polar_multi}
\end{minipage}
\hfill%
\begin{minipage}[c]{.32\linewidth}
  \centering
  \includegraphics[width=\linewidth]{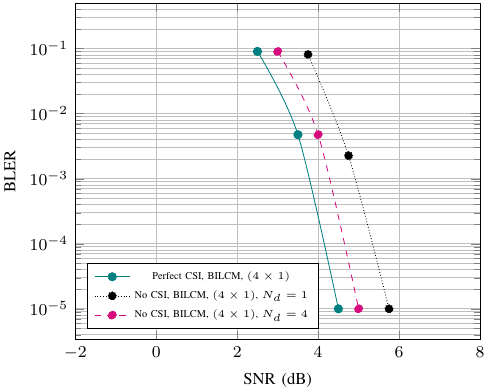}
  \caption{Block Error Rate, 48 bits(TBs+CRC), NRLDPC BICM, R=48/64, Layered belief propagation decoder, iteration=30, QPSK modulation, 1 OFDM symbol, 4 PRBs, 48 REs (32 data, 16 DMRS), $N_T$=1, $N_R$ = 4,  $\alpha=1$, LOS channel.}
  \label{fig:bicm_pucch48bits_LDPC_multi}
\end{minipage}
\begin{minipage}[c]{.32\linewidth}
  \centering
  \includegraphics[width=\linewidth]{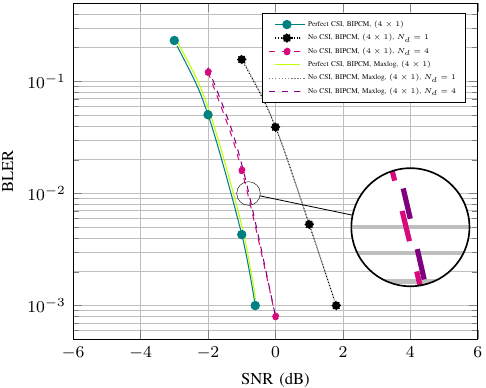}
  \caption{Block Error Rate, 48 bits(TBs+CRC), NRPOLAR BICM, R=48/64, (CRC)-aided successive-cancellation list decoder(List length=8), QPSK modulation, 1 OFDM symbol, 4 PRBs, 48 REs (32 data, 16 DMRS), $N_T$=1, $N_R$ = 4, $\alpha=1$, LOS channel, Accurate Metric vs Max-log metric. }
  \label{fig:bicm_pucch48bits_polar_multi_max_log}
\end{minipage}
\hfill%
\end{figure*}

The simulations are based on NR POLAR and NR LDPC coding schemes paired with QPSK modulation. The transmission process involves a transport block length of 48 bits. The resource population process is conducted using a single OFDM symbol with 4 PRBs and 48 REs (32 REs for data components and 16 REs for DMRS components), wherein the DMRS sequences occupy 4 REs per PRB.
The results illustrated in Figure~\ref{fig:bicm_pucch48bits_polar_multi} show the performance of the Bit Interleaved Coded Modulation (BICM) for joint estimation and detection over a  LOS channel, specifically when $\alpha=1$ is  assessed to understand the performance discrepancy between the \textit{Perfect CSI} and \textit{No CSI} situations in extreme coverage scenarios characterized by low signal-to-noise ratio.
Note that the $N_d =1$ case also  refers to the conventional receiver utilizing separate channel estimation.
 The joint estimation/detection($N_d=4$) approach yields  a perfomance gain of  $1.25$ dB, $1.5$ dB and $1.75$ fB over 2, 4 and 8 receive antennas respectively.
From this insight, it is apparent  that when the number of antennas increases, the performance gap between the \textit{Perfect CSI} and the \textit{No CSI} situations (e.g., $N_d =4$) expands.
Similarly, the results in Figure~\ref{fig:bicm_pucch48bits_polar_multi} using BIPCM are congruent with those presented in Figure~\ref{fig:bicm_pucch48bits_LDPC_multi} that employs BILCM, in both single and multiple antenna configurations. Although the code rates and transmission parameters are identical, BIPCM offers significantly better performance gains than BILCM. This is potentially due to the fact that the 3GPP polar code has been optimized for very short block lengths, while the 3GPP LDPC code targets much longer transport block lengths.

Furthermore, Figure \ref{fig:bicm_pucch48bits_polar_multi_max_log} indicate that the max-log metric performs nearly as well as the accurate metric.
This leads to the conclusion that  when Gray-mapped constellations are employed, the max-log metric is known to have a minimal impact on receiver performance. However, as the modulation order increases, the difference in performance between optimal and suboptimal techniques for generating LLRs becomes significant as discussed  in \cite{Szczecinski2002}. The logarithmic calculations tied to the precise metric  add an extra layer of complexity when incorporating the requisite multiplicative and additive operations during LLR processing.

\begin{figure*}[!ht]
  \begin{minipage}[c]{.32\linewidth}
    \centering
    \includegraphics[width=\linewidth]{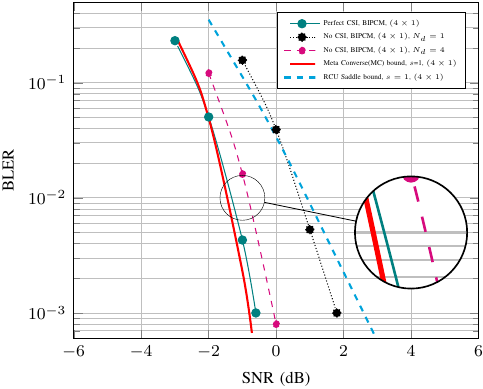}
    \caption{Block Error Rate, 48 bits(TBs+CRC), NR POLAR BICM, R=48/64, (CRC)-aided successive-cancellation list decoder(List length=8), QPSK modulation, 1 OFDM symbol, 4 PRBs, 48 REs (32 data, 16 DMRS), $N_T$=1, $N_R$ = 4 over  LOS channel, vs converse and achievability bounds.}
    \label{fig:bicm_pucch48bits_polar_multi_bound}
  \end{minipage}
  \hfill%
  \begin{minipage}[c]{.32\linewidth}
    \centering
    \includegraphics[width=\linewidth]{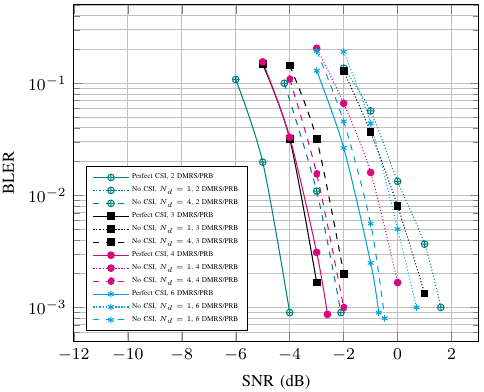}
    \caption{Block Error Rate, 24 bits(TBs), NR POLAR BICM, (CRC)-aided successive-cancellation list decoder, QPSK modulation, 1 OFDM symbol, 4 PRBs, 48 REs, set of  ($\{24, 32, 36, 40\}$ data, $\{24, 16,12, 8\}$ DMRS), $N_T$=1, $N_R$ = 4, $\alpha=1$, LOS channel.}
     \label{fig:DMRS_density_4rx}
  \end{minipage}
    \hfill%
  \begin{minipage}[c]{.32\linewidth}
    \centering
    \includegraphics[width=\linewidth]{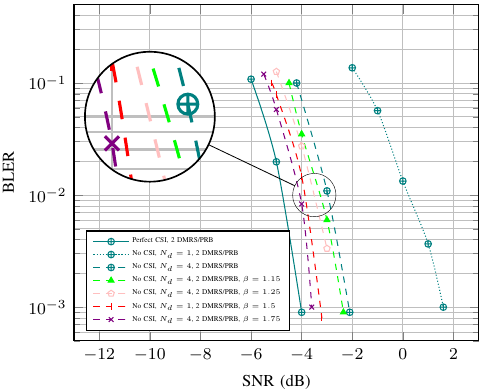}
    \caption{Block Error Rate, 24 bits(TBs), NR POLAR BICM, (CRC)-aided successive-cancellation list decoder, QPSK modulation, 1 OFDM symbol, 4 PRBs, 48 REs, ($40$REs= data, $8$REs= DMRS), $N_T$=1, $N_R$ = 4, $\alpha=1$, LOS channel, adaptive power adjustment via $\beta$ with $N_d = 4$ and $N_d =1$. }
    \label{fig:adaptative_power_ajustement}
  \end{minipage}

  \end{figure*}

Finally, we can assess the above results with respect to the finite block length bounds that have been established in the scientific literature \cite{Polyanskiy2010, Martinez2011, Erseghe2016, Ostman2019jrnal, Xhemrishi2019}.
 For a more comprehensive understanding of the bounds utilized in Figure \ref{fig:bicm_pucch48bits_polar_multi_bound}, interested readers are encouraged to refer to the works of authors \cite{Xhemrishi2019, Martinez2011}. For this purpose, we consider  the {\em metaconverse} (MC) and {\em Random Coding Union} (RCU) bounds for a thorough comparative analysis.
  It can be observed that when the block error rate (BLER) reaches a threshold of $1\%$, the performance difference between the \emph{MC Bound} and the  \emph{No-CSI}($N_d =4$) is $0.7$ dB, compared to $2.2$ dB for the  \emph{No-CSI}($N_d =1$).

\subsection{Impact of DMRS density}
In instances where the reference and data symbols are jointly conveyed in common OFDM symbols, we can look into the impact of DMRS density on performance.
Figure \ref{fig:DMRS_density_4rx} depicts the performance on the LOS channel in the situations of \emph{Perfect CSI} and \emph{No CSI ($N_d =1$, $N_d =4$)} depending on the density of DMRS per PRB.
In essence, fewer DMRS has merit of additional coding rates. Therefore,  performance improves as DMRS density decreases. However, it should be noted  that even with $N_d =4$, a low DMRS density setup expands the performance gap between \textit{Perfect CSI}  and \textit{No CSI}.
It may be advantageous in some instances to maintain the density of DMRSs in a certain sweet spot or simply to rely on sparse or even low DMRS density while increasing their power via an adaptive adjustment. More specifically, a precise approach is to identify the configuration with the minimum number of DMRSs which allows the transmitter to slightly increase the power of the underlying signals. However, choosing a low DMRS density has a detrimental effect on channel estimate quality. Even if the receiver with block detection ($N_d =4$) seems to be less sensitive to it  with respect to the conventional receiver ($N_d =1$). There appears to be a sweet spot in terms of DMRS density per PRB, as evidenced by the results presented in Figure~\ref{fig:DMRS_density_4rx}. Therefore, the ideal DMRS distribution setup is obtained by incorporating four DMRSs per {\em physical resource block} compared to those employing two, three or six DMRSs per PRB , using the block detection principle ($N_d =4$). In practice, transmission with a low density of DMRS appears to be more valuable.
Consequently, it is advisable to consider configurations with either one or two DMRS per PRB. However in order to reap from the low DMRS density, it is important to carry out some sort of \emph{adaptive  DMRS/data power adjustment} that would enhance the channel estimate accuracy, leading to an  improvement in performance from a holistic perspective. For this purpose, the system model can be reconceived as $\mathbf y_i=\left( \mathbf x^{(d)} + \beta\mathbf x^{(p)}\right)\mathbf h_i + \mathbf z_i$. The adaptive power adjustment procedure is contingent on the values of $\beta$. The DMRS Power is to be slightly increased in a judicious fashion since $\beta$ must be perfectly calibrated to ensure compliance with potential radio frequency constraints.

As depicted in Figure ~\ref{fig:adaptative_power_ajustement}, the performance improvement can be observed as a function of varying values of $\beta$. The optimal performance enhancement is achieved when $\beta$ is set to $1.75$. It is noted that by selecting $\beta = 1.5$, a gain of $1$ dB can be attained when $N_d=4$.

 Overall, the implications of varying DMRS density  within the 3GPP standard are significant. Specifically, it is feasible to reduce the number of DMRS per PRB to one or two, while allowing the {\em User Equipment} (UE) to adjust the power allocation between the DMRS and data transmission. This flexibility in DMRS density and power allocation is transparent to the receiver.
 \balance
 \section{Conclusions}
This paper presented novel bit-interleaved coded modulation metrics for joint estimation detection using a training or reference signal transmission strategy for short to long block length channels.
The proposed techniques take advantage of joint estimation/detection.
We considered transmissions where reference signals are interleaved with data and both are transmitted over a small number of OFDM symbols so that near-perfect channel estimation cannot be achieved.
Our findings demonstrate that BICM metrics combined with the joint estimation/detection principle can be used to achieve detection performance that is close to that of a coherent receiver with perfect channel state information for both polar and LDPC coded configurations. Furthermore, we showed that for transmissions with low DMRS density, a good trade-off can be achieved in terms of additional coding gain and improved channel estimation quality by adaptive DMRS power adjustment.
\section*{Acknowledgment}
This work has been supported by Qualcomm.
 \appendix
 \label{appendix:llrfading}
 \renewcommand{\thesection}{\Alph{section}.\arabic{subsection}}
 \setcounter{section}{0}
 Since $\theta_i$ is unknown and randomly distributed over $[0, 2\pi)$, the conditional probability density function can be written as follows:
 \begin{align}
 \begin{array}{r}
 p\left(\mathbf{y}_i \mid \mathbf{x}\right)
 =\displaystyle\frac{1}{2 \pi \operatorname{det} \Phi} \displaystyle\int_0^{2 \pi}
 \exp \left(-\frac{1}{2}\left(\mathbf{y}_i-\ \mu\{\mathbf x, \theta_i\} \right)^{\dagger}\right. \\
 \left.\Phi^{-1}\left(\mathbf{y}_i-\mu\{\mathbf x, \theta_i\}\right)\right) \mathrm{d} \theta_i
 \end{array}
 \end{align}
Saying  $\mu\{\mathbf x, \theta_i\}=  \sqrt{\alpha} e^{j \theta_i}\mathbf x$.
\subsubsection*{\bf \small{Covariance Matrix}}~\\ Knowing that
 \begin{equation}
   \mathbf y_i - \sqrt{\alpha} e^{j \theta_i} \mathbf x = \sqrt{1-\alpha} \mathrm h_{i,f}\mathbf x + \mathbf z_i,
 \end{equation}

 \begin{equation}
 \begin{aligned}
 \Phi&\triangleq
 \frac{1}{2}\mathbb{E}\left[\left(\sqrt{1-\alpha} \mathrm h_i^{(f)}\mathbf x + \mathbf z_i \right)\left(\sqrt{1-\alpha}\mathrm h_i^{(f)}\mathbf x + \mathbf z_i\right)^\dag\right]\\
 &\triangleq  (1-\alpha)  \mathbf x\mathbf x^\dag \sigma^2_h + \sigma^2_z \mathbf I_N, \text{where } \sigma^2_h =1\\
 \end{aligned}
 \end{equation}

 \subsubsection*{\bf \small{Determinant}}
 \begin{equation}
 \begin{aligned}
 \operatorname{det} \Phi &= \operatorname{det}\left( (1-\alpha)  \mathbf x\mathbf x^\dag + \sigma^2_z \mathbf I \right)
  = \operatorname{det}\left( \sigma^2_z \mathbf I + (1-\alpha)  \mathbf x\mathbf x^\dag \right)\\&=c\left(\mathrm{N_0}+{2(1-\alpha)\|\mathbf{x}\|^2}\right), \text{ where  $c=\displaystyle\frac{1}{2}\left(\frac{\mathrm{~N}_0}{2}\right)^{N-1}$},
 \\&\propto \mathrm{N_0}+{2(1-\alpha)\|\mathbf{x}\|^2}.
 \end{aligned}
 \end{equation}

 \subsubsection*{\bf  \small{Inverse of $\Phi$ }} {\em The Woodbury matrix identity} \cite{Woodbury1950}
\begin{equation}
(\mathbf A+\mathbf U \mathbf C \mathbf V)^{-1}=\mathbf A^{-1}-\mathbf A^{-1} \mathbf U\left(\mathbf C^{-1}+\mathbf{V A}^{-1} \mathbf U\right)^{-1} \mathbf {V A}^{-1},
\end{equation}
 where $\mathbf A$, $\mathbf U$, $\mathbf C$ and $\mathbf V$ are conformable matrices: $A$ is $n\times n$, $\mathbf C$ is $k\times k$, $\mathbf U$ is $n\times k$, and $\mathbf V$ is $k\times n$.
%

 Let's say :$
 \left\{\begin{array}{l}
  \displaystyle \mathbf A= \sigma^2_z \mathbf I, \quad  \displaystyle \mathbf C =(1-\alpha)\mathbf I \\
   \displaystyle \mathbf U= \mathbf x,  \quad \displaystyle \mathbf V= \mathbf x^\dag
 \end{array}\right.$

 \begin{equation}
 \begin{aligned}
 \Phi^{-1}
   =\frac{2}{N_0}- \frac{2}{N_0} \mathbf x
   \left(\frac{2(1-\alpha) }{N_0 +2(1-\alpha) \left\|\mathbf x \right\|^2} \right)\mathbf x^\dag
 \end{aligned}
 \end{equation}
 let's say  $\beta_x = \frac{2(1-\alpha) }{N_0(N_0 +2(1-\alpha) \left\|\mathbf x \right\|^2)}$, then $ \Phi^{-1}=\frac{2}{N_0}- 2 \mathbf x
  \beta_x\mathbf x^\dag$
  \vspace{-0.5em}
 \subsubsection*{\bf \small{Likelihood function}}
\begin{equation}
\begin{aligned}
\mathrm q\left(\mathbf{x},\mathbf y_i\right)&
=\displaystyle \frac{1}{2 \pi \operatorname{det} \Phi}\displaystyle
 \int_{0}^{2\pi}\exp \left(-\frac{1}{\mathrm{N_0}}\left\|\mathbf y_i- \sqrt{\alpha} e^{j \theta_i}\mathbf x\right\|^2 \right. \\&
\left.+ \beta_x \left\|\left(\mathbf y^\dag_i- \sqrt{\alpha} e^{-j \theta_i}\mathbf x^\dag\right)\mathbf x\right\|^2\right)\mathrm{d} \theta_i.
\end{aligned}
\end{equation}
By extending the terms into the exponential, ignoring those that are independent of $\mathbf x$,
 the likelihood function is equal to
 \begin{equation}
 \begin{array}{r}
 q\left(\mathbf{x},\mathbf y_i\right)
 =\displaystyle \frac{1}{2 \pi \operatorname{det} \Phi}\exp \left(- \alpha \left\|\mathbf x \right\|^2 \left(\frac{1}{N_0} - \beta_x \left\| \mathbf x\right\|^2\right) \right. \\
 \left.
 \beta_x \left| \mathbf x^\dag \mathbf y_i \right|^2\right) \displaystyle \int_{0}^{2\pi}\exp \left( 2\sqrt{\alpha} \left(\frac{1}{N_0} - \beta_x \left\| \mathbf x\right\|^2\right)\right. \\
 \left.|
   \mathbf x^\dag\mathbf y_i|\mathrm{cos}\left({\phi_i} + \theta_i \right)\right)\mathrm{d} \theta_i \nonumber
 \end{array}
 \end{equation}
 knowing that $\displaystyle \frac{1}{\pi}\int_{{\varphi}=0}^{\pi}\exp(zcos(\varphi))\mathrm{d} \varphi=\operatorname{I_0(z)} $ \cite{Gradshteyn95}.\\
 Saying  $\mathbf L_x = N_0+ 2(1-\alpha)  \left\|\mathbf x \right\|^2$, and then after ignoring multiplicative term that are independent of $\mathbf x$, it comes
  \begin{equation}\label{eqn:llrfunction_proof}
  \begin{aligned}
  &q\left(\mathbf{x},\mathbf y_i\right)\propto \frac{1}{\mathbf L_x}\exp \left(- \alpha \left\|\mathbf x \right\|^2\left(\frac{1}{N_0} - \mathbf \beta_x \left\| \mathbf x\right\|^2\right) \right. \\& \left.  + \mathbf \beta_x \left| \mathbf x^\dag \mathbf y_i \right|^2\right)\times\operatorname{I_0}\left( 2\sqrt{\alpha} \left(\frac{1}{N_0} - \mathbf \beta_x \left\| \mathbf x\right\|^2\right)\left|\mathbf x^\dag\mathbf y_i\right|\right)
  \end{aligned}
  \end{equation}
  Expressing $\mathbf \beta_x$  w.r.t  $\mathbf L_x$  $\rightarrow$
  $\mathbf \beta_x =\frac{1}{\left\|\mathbf x \right\|^2 N_0} - \frac{1}{\left\|\mathbf x \right\|^2\mathbf L_x}$
 \ifCLASSOPTIONcaptionsoff
   \newpage
 \fi


\vspace{12pt}

\end{document}